\documentclass[12pt]{iopart}
\usepackage{bm}
\usepackage{amssymb}
\usepackage{epsfig}
\usepackage{float}
\newtheorem{theorem}{Theorem}
\newtheorem{corrolary}{Corollary}
\begin{document}

\newcommand{\bk}{{\bf k}}
\newcommand{\blm}{{\bf m}}
\newcommand{\bn}{{\bf n}}
\newcommand{\bl}{{\mathbf l}}
\newcommand{\bls}{{\mathbf s}}

\title[Quantum de Moivre-Laplace theorem]{Quantum de Moivre-Laplace theorem for noninteracting indistinguishable  particles in random networks}

\author{V. S. Shchesnovich}

\address{Centro de Ci\^encias Naturais e Humanas, Universidade Federal do
ABC, Santo Andr\'e,  SP, 09210-580 Brazil }

\begin{abstract}
The asymptotic form of  the  average probability  to count  $N$ indistinguishable identical particles  in a small number $r \ll N$ of binned-together output ports  of a  $M$-port Haar-random unitary network, proposed recently  in  \textit{Scientific Reports} \textbf{7}, 31 (2017)  in a  heuristic manner with some numerical confirmation,  is presented  with the  mathematical rigor  and generalized to an arbitrary (mixed) input state of $N$ indistinguishable particles.   It is shown that, both in the classical (distinguishable particles) and quantum (indistinguishable particles) cases, the average counting probability  into $r$ output bins    factorizes into  a product of $r-1$ counting probabilities into two bins. This fact relates the  asymptotic  Gaussian law  to the de Moivre-Laplace theorem in the classical case and similarly in the quantum case where an analogous theorem  can be stated.  The results have applications to  the setups where  randomness plays a  key role, such as  the multiphoton propagation in disordered media and the scattershot  Boson Sampling. 
\end{abstract}
%Uncomment for PACS numbers title message
%\pacs{00.00, 20.00, 42.10}

\maketitle

\section{Introduction}
\label{sec1}

Recently it was argued   \cite{GL} that the probability to count $N$ indistinguishable particles, bosons or fermions,  in binned-together output ports of a unitary $M$-port, averaged over the   Haar-random unitary matrix representing the multiport or, for a fixed multiport, over the input configurations of the particles,  takes  the asymptotic Gaussian form   as $N\to  \infty$ with  the particle density  $N/M$  being constant. The  quantum statistics of bosons or fermions enters  the Gaussian law precisely  through  the particle density. In this respect, the random multiport with identical particles at its input  can be thought of as a quantum variant of the  Galton board   (invented to expose the convergence of the multinomial distribution to a Gaussian one) for  the  indistinguishable identical particles correlated due to their quantum statistics. 

 The  quantum asymptotic  Gaussian law generalizes to the quantum-correlated particles   the  well-known asymptotic result    for the multinomial distribution, originally due to works of A.~de Moivre, J.-L.~Lagrange, and P. S.~Laplace (for a  historical review, see Ref. \cite{Hald}). The multinomial distribution applies  when the  identical particles are sent one at a time at the input (i.e., they are  distinguishable particles).  The  asymptotic  Gaussian law  exposes the effect  of the quantum statistics of identical particles on their behavior,   explored previously \cite{E5,MCMS,BB,StatBench},  applicable to the setups where  randomness plays a  key role, such as  the multiphoton propagation through disordered media \cite{DM,QS2P,PNCDM}. It is known that   the  quantum interference may  result in  the common forbidden  events  for bosons and fermions in some  special (symmetric) multiports  \cite{MPI,FourExp}, obscuring the role played by the quantum statistics.    It should be stressed that  the   complexity of behavior  of bosonic particles in  linear unitary networks  asymptotically challenges  the digital computers, which is  the essence of  the Boson Sampling idea  \cite{AA,SCBS}  with the  proof-of-principle  experiments \cite{E5,E1,E2,E3,E4,ULO,E6,E7}.  The Gaussian law is applicable to the scattershot Boson Sampling \cite{SCBS,E6}, where  randomness in the setup is due to the heralded photon generation in random input ports (see also section \ref{sec2} below).  
 
The main purpose of the present work is to give a mathematically rigorous  derivation of the asymptotic Gaussian law previously but  heuristically proposed before \cite{GL} with some numerical confirmation. The main technical tool in the proof is the discovered factorization of the average $r$-bin counting probability distribution  as a series of layered  probability distributions for the binary case (two bins).   For instance, this fact is used to show the equivalence of the classical asymptotic Gaussian law for the $r$-bin partition to the  de Moivre-Laplace  theorem \cite{Hald}.  This equivalence extends to  the quantum case as well, suggesting  the interpretation of the respective asymptotic Gaussian law as a quantum version  of the de Moivre-Laplace theorem, where the events (particle counts in this case) are quantum-correlated due to the indistinguishability of the particles.

Section \ref{sec2} contains a brief statement of the problem  and a rigorous formulation of the main results in the form of two theorems.  The theorems are proven in section \ref{sec3}, where the binary classical case is briefly reminded in section \ref{sec3A} and the $r$-bin case is considered in section \ref{sec3B}. The $r$-bin quantum case is analyzed    in section \ref{sec3C}, where similarities of the classical and quantum cases are highlited. Appendices A, B, and C contain mathematical details of the proof.  In section \ref{sec4}  the results in theorems 1 and 2 are generalized to \textit{arbitrary} (mixed)  input state of indistinguishable particles. Finally, in section \ref{sec5} a brief summary  of the results is given.

%%%%%%%%%% SEC 2 %%%%%%%%%%%%%%%%%%
\section{The counting probability of identical particles in a random  multiport with binned-together output ports}
\label{sec2}

\begin{figure}[htb]
\begin{center}
\includegraphics[width=0.5\textwidth]{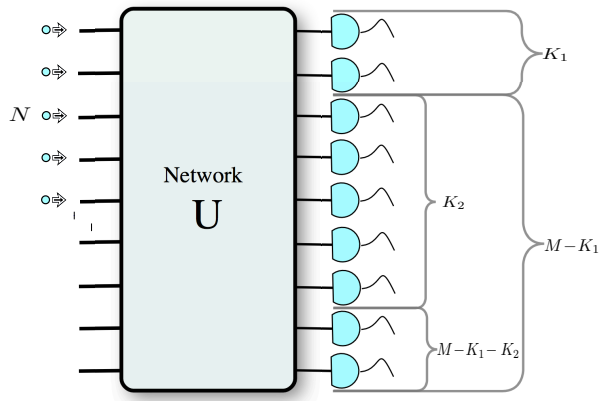}
\caption{A quantum network, having a unitary matrix $U$,  with $N$ indistinguishable   identical particles at its  input and  binned-together output ports (two or more bosons as well as  classical particles  may share the same input port). The braces illustrate the successive    binary partition into $r$ bins ($r=3$  in this case).  We are interested  in the probability of counting   $\mathbf{n} = (n_1,n_2,n_3)$  particles  in the output  bins with $K_1,K_2,K_3$, where $K_3 = M - K_1-K_2$.  } \label{F1}
\end{center}
\end{figure}

Consider   a  unitary quantum $M$-port  network (i.e., where there are $M$ input and output ports) described by  the unitary matrix $U$  connecting the input $|k,in\rangle$ and output $|l,out\rangle$   states as follows  $|k,in\rangle = \sum_{l=1}^M U_{kl}|l,out\rangle$ and  whose output ports  are partitioned into $r$ bins having   $\mathbf{K}\equiv (K_1,\ldots,K_r)$ ports. We are interested in the  probability of counting  $N$ noninteracting  identical  particles, impinging at the network input, in the binned-together output ports, as in Fig. \ref{F1}  (for more details  see also Ref. \cite{GL}). We are interested in the  average probability in a random unitary multiport  (except where  stated otherwise, here and below  the term  ``average'' means  the average over the Haar-random unitary matrix $U$; where necessary we will use  the  notation $\langle \ldots \rangle$). A random unitary optical multiport  can be   experimentally realized  with a very high fidelity \cite{ULO} and without explicit  matrix calculations   \cite{DDHRU}.  We are interested in the  probability in binned-together output ports since in the quantum case the average probability of an output configuration of indistinguishable bosons is exponentially small and, therefore, hard to estimate experimentally (see below).

Let us consider first distinguishable particles. 
%%FIG 1
The term ``distinguishable particles"   applies here to quantum particles  in different states with respect to the degrees of freedom not affected by a multiport \cite{HOM,Ou,VS,Tichy},  such as the  arrival time  in the case of  photons (e.g., particles  sent one at a time through the multiport; however, the time-resolving detection scheme \cite{Tamma1,Tamma2}  unitarily mixes also the internal degrees of freedom, making them  the operating modes).   The  average probability for a single  particle  from  input port $k$ to land into  bin $i$   reads \cite{GL} $p_i = \sum_{l\in K_i} \langle |U_{k,l}|^2\rangle =  q_i \equiv K_i/M$, since $\langle  |U_{kl}|^2\rangle  = 1/M$, where $|U_{k,l}|^2$ is the   probability of the transition $k\to l$.    For identical  particles sent one at  a time through a  random multiport, the average probability to count $\mathbf{n}\equiv (n_1,\ldots,n_r)$ particles in the output bins becomes   a multinomial distribution
\begin{equation}
P^{(D)}(\mathbf{n}|\mathbf{K}) = \frac{N!}{\prod_{i=1}^rn_i!}\prod_{i=1}^rq_i^{n_i}.
\label{E1}\end{equation}
Under the above conditions, Eq. (\ref{E1}) applies also for a fixed unitary multiport with the  averaging performed over the uniformly random input ports of the particles   \cite{GL}.

Now let us   consider indistinguishable particles  (assuming in the case of fermions up to  one particle  per network port, as in Fig. \ref{F1}).   For the  input   $\mathbf{k}= (k_1,\ldots,k_N$) and  output $\mathbf{l}= (l_1,\ldots,l_N$)  configurations the average probability  of the transition $\mathbf{k}\to \mathbf{l}$  is just the inverse of the number of Fock states of $N$ bosons (fermions) in $M$ ports: $ p^{(B)}(\mathbf{l}|\mathbf{k})= \frac{N!}{(M+N-1)\ldots M}$ $\bigl( p^{(F)}(\mathbf{l}|\mathbf{k})  = \frac{N!}{M \ldots (M-N +1)}\bigr)$. This observation leads to  the following quantum equivalent of Eq. (\ref{E1}) (with the upper signs for bosons and the lower ones  for fermions)  \cite{GL}
\begin{eqnarray}
\fl \qquad P^{(B,F)}(\mathbf{n}|\mathbf{K})  &=&\frac{N!}{(M\pm N\mp1)\ldots M}\prod_{i=1}^r\frac{(K_i\pm n_i\mp 1)\ldots K_i}{n_i!} \nonumber\\
\fl &=&
 P^{(D)}(\mathbf{n}|\mathbf{K}) \frac{\prod_{i=1}^r\left(\prod_{l=0}^{n_i-1}\left[1\pm l/K_i\right]\right)}{\prod_{l=0}^{N-1}\left[1\pm l/M\right]}\equiv P^{(D)}(\mathbf{n}|\mathbf{K})Q^{(\pm)}(\mathbf{n}|\mathbf{K}).
\label{E2}\end{eqnarray}
As in the classical case, the  probability formula (\ref{E2}) applies also to a fixed unitary multiport with the  averaging performed uniformly over the  input configurations  $\mathbf{k}$  allowed by the quantum statistics \cite{GL}. Moreover, as shown in section \ref{sec4} below, the average probability in Eq. (\ref{E2}) actually applies to \textit{arbitrary} mixed input state of $N$ indistinguishable particles. 

Now, our  focus on the binned-together output ports  for a large multiport and large number of particles can be explained. Assuming that only asymptotically  polynomial in $N$ number of experimental runs is accessible (due to the decoherence  or as in  verification protocols for the Boson Sampling \cite{E5,BB,AA,E6,Drumm}), an exponentially small in $N$ probability cannot be estimated.  Assuming scaling up for a fixed particle density  $\alpha=  N/M$, the probability of a particular configuration of indistinguishable bosons  at the output of a random $M$-port, on average, is asymptotically exponentially small in $N$ (see also Ref. \cite{Drumm,Arkhipov}): 
\begin{equation}
\fl p^{(B)}(\mathbf{l}|\mathbf{k}) =   \sqrt{2\pi N(1+\alpha)}e^{-\gamma N }\left[1+ \mathcal{O}\left(\frac{1}{N}\right)\right],\;
      \gamma =  \ln\left(\frac{1}{\alpha}\right)+ \left(1\pm \frac{1}{\alpha}\right) \ln\left(1\pm \alpha\right).
\end{equation}
Though the density $\alpha$ is not fixed below (in the statement of theorem 1),  it is natural to consider the asymptotic limit at a fixed density, i.e.,  when both  the number of particles and the number of ports tend  to infinity (for bosons, there is also the high-density case $M = \mathcal{O}(1)$ as $N\to \infty$, see Corollary 1 below).

Eqs. (\ref{E1}) and (\ref{E2}) are  good approximations to the average counting probability in the binned-together output modes for  $N$  identical particles in  the  disordered media \cite{DM,QS2P,PNCDM}, chaotic cavities \cite{MCMS}. It also can be applied to   the scattershot Boson Sampling \cite{SCBS,E7} (due to the uniform averaging over the input configurations with up to one particle per input port in the low density limit $M \gg N^2$, when  the contribution from the bunched configurations  scales  as $\mathcal{O}(N^2/M)$ \cite{Arkhipov}).  In such setups, Eq. (\ref{E1}) is not, however, the   exact average probability for distinguishable (classical) particles, since for  $N$ simultaneous distinguishable particles  at the input there is  an extra factor \cite{GL}  due to  the correlations between the matrix elements $|U_{kl}|^2$.

In the proof of the asymptotic Gaussian law of Ref. \cite{GL}  we will use that  the $r$-bin case can be considered as a set of  layered  $r-1$   binary cases, as illustrated in Fig. \ref{F1} to the second layer. Indeed, both in the classical  and quantum cases there is  an  exact  factorization  of the average counting probability into binned-together output ports  (see   \ref{appA}):
\begin{eqnarray}
P(\mathbf{n}|\mathbf{K})  &= & P(n_1,N_1-n_1|K_1,M_1-K_1)P(n_2,N_2-n_2|K_2,M_2-K_2)\nonumber\\
&\ldots &P(n_{r-1},N_{r-1}-n_{r-1}|K_{r-1},M_{r-1}- K_{r-1}),
\label{E3}
\end{eqnarray}
where  $N_1=N$, $M_1=M$ and for $s=2,\ldots,r-1$
\begin{eqnarray}
N_s = N - \sum_{i=1}^{s-1}n_i,\quad M_s = M - \sum_{i=1}^{s-1}K_i.
\label{E4}\end{eqnarray}
Eq. (\ref{E3}) shows the key role played by the binary case, for which we will use also another notation $P_{N,M}(n|K) = P(n,N-n|K,M-K)$, with the independent  variables as the arguments. Below we will need the following definitions (for $s\ge 2$):
\begin{equation}
\bar{q}_s = \frac{K_s}{M_s} = \frac{q_s}{1-\sum_{i=1}^{s-1}q_i},\quad \bar{x}_s = \frac{n_s}{N_s} = \frac{x_s}{1-\sum_{i=1}^{s-1}x_i}, \quad x_i = \frac{n_i}{N},
\label{E5}\end{equation}
where $0\le \bar{q}_s\le 1$ takes the place of $q_s$ in  the $s$th layer of the binary partition of the average classical  probability for  the $r$-bin case in   Eq. (\ref{E3}), i.e., $P^{(D)}_{N_s,M_s}(n_s|K_s) = \frac{N_s!}{n_s! (N_s-n_s)!} {\bar{q}_s}^{\,n_s}(1-\bar{q}_s)^{N_s-n_s}$. Let us differentiate by  $\sigma$  the three cases of identical particles, where  bosons correspond to $\sigma = +$, fermions to $\sigma =-$, and distinguishable particles to $\sigma = 0$.  In the case when $Y$ is of order $X$, i.e, when  there is such $C>0$ (independent of $X$)  that $Y\le C X$,  we use the notation $Y = \mathcal{O}(X)$. The following two theorems state the main results.

\begin{theorem}
Consider   the Haar-random unitary $M$-port with the   binned together output ports into $r$  sets of  $K_1,\ldots,K_r$ ports. Then, as $N,M\to \infty$ for a fixed $q_i = K_i/M>0$, the average probability to count  $\mathbf{n}=(n_1,\ldots,n_r)$ identical particles into the $r$ bins such that  
\begin{equation}
|n_i - N q_i| \le AN^{\frac23-\epsilon},\quad A>0,\quad 0<\epsilon< \frac16
\label{E6}\end{equation}    has the following asymptotic  form   
\begin{equation}
\fl \quad
P^{(\sigma)}(\mathbf{n}|\mathbf{K}) = \frac{\exp\left\{-N\sum_{i=1}^r \frac{(x_i-q_i)^2}{2(1+\sigma\alpha)q_i} \right\} }{\left( 2\pi [1+\sigma \alpha]N\right)^{\frac{r-1}{2}} \prod_{i=1}^r\sqrt{q_i}}\left\{1+ \mathcal{O}\left(\frac{(1-\alpha\delta_{\sigma,-})^{-3} }{N^{3\epsilon}}+\frac{\alpha\delta_{\sigma,+}}{N}\right)\right\}.
\label{E7}\end{equation}
\end{theorem}
An important note is in order. In the course of the proof (see section \ref{sec3}) it is also established that the $r$-bin asymptotic Gaussian on the right hand side of Eq. (\ref{E7}) satisfies the same factorization as the average counting probability, Eq. (\ref{E3}), to the error of the asymptotic approximation.

For a finite density $\alpha$, in the quantum case  the  error in Eq. (\ref{E7})  scales as $ \mathcal{O}(N^{-3\epsilon})$. In this case  we get   $x_i = q_i$  in  the limit $N\to \infty$ for  bosons, fermions, and distinguishable particles.   In the  usual presentation of the  classical result  $\epsilon = 1/6$ \cite{Gnedenko}, with this choice the error in  Eq. (\ref{E7}) scales as $\mathcal{O}(N^{-\frac12})$ (this choice, however,  invalidates the error estimate in theorem 2,  Eq. (\ref{E8}) below, thus it is not allowed).

 Since not all particle counts are covered by Eq. (\ref{E6}), theorem 1 does not guarantee the asymptotic Gaussian to be an uniform approximation for all $\mathbf{n}$.  However, if Eq. (\ref{E6}) is violated,  then the respective particle counts occur with an exponentially small  probability (asymptotically undetectable in an experiment with only a polynomial in $N$ number of runs). This is stated in the following theorem. 
\begin{theorem}
 The average probability of the particle counts  $\mathbf{n}$ violating    Eq. (\ref{E6}) for $N,M\to \infty$ and  $\alpha= N/M$  being fixed satisfies 
 \begin{equation}
P^{(\sigma)}(\mathbf{n}|\mathbf{K}) = \mathcal{O}\left(N^{s(\sigma)} \exp\left\{-\frac{A^2}{1+\sigma\alpha} N^{\frac13-2\epsilon}\right\}\right),
\label{E8}\end{equation}
where  $A$ is from Eq. (\ref{E6}), whereas $s(0) = 1/2$  (distinguishable particles),  $s(+) = 1/2$ (bosons) and  $s(-) = 5/2$ (fermions).  
\end{theorem}

In Ref. \cite{GL} the  high-density limit for bosons was mentioned,  realized for $N\to\infty$  and  $M = \mathcal{O}(1)$. This case is a corollary to theorem 1. 
\begin{corrolary}
As $N\to \infty$  and a fixed $M \gg 1$ the average probability to count $\mathbf{n}=(n_1,\ldots,n_r)$ identical particles into $r$ bins with  $K_1,\ldots,K_r$ output ports of a Haar-random unitary $M$-port, such that Eq. (\ref{E6}) being satisfied,    has the following approximate  asymptotic  form   
 
\begin{equation}
\! \! \! \! \! \! \!\!\!\! \! \! \! \! \!  
P^{(B)}(\mathbf{n}|\mathbf{K}) = \frac{M^{\frac{r-1}{2}}\exp\left\{-M\sum_{i=1}^r \frac{(x_i-q_i)^2}{2q_i} \right\} }{\left( 2\pi N^2\right)^{\frac{r-1}{2}} \prod_{i=1}^r\sqrt{q_i}}\left\{1+ \mathcal{O}\left(\frac{1}{M} + \frac{1}{N^{3\epsilon}}\right) \right\}. 
\label{E9}\end{equation}
\end{corrolary}
Corollary 1 tells us that in the limit $N\to \infty$ the relative  particle counting variables  $x_1,\ldots,x_r$ are   approximated  by the continuous Gaussian random variables (similar as in  the classical case in Ref. \cite{Gnedenko}): 
\begin{equation}
x_i = q_i + \frac{\xi_i}{\sqrt{M}},
\label{E10}\end{equation}
where $\xi_1,\ldots,\xi_r$  are random variables satisfying the constraint $\sum_{i=1}^r\xi_i = 0$ with a  Gaussian joint probability density 
\begin{equation}
\rho = \frac{\exp\{-\sum_{i=1}^r \frac{\xi_i^2}{2q_i}\} }{\left(2\pi\right)^{\frac{r-1}{2}} \prod_{i=1}^r \sqrt{q_i}}.
\label{E11}\end{equation}
(The factor $\left(\frac{M}{N^2}\right)^{\frac{r-1}{2}}$ In Eq. (\ref{E9}) allows to convert  the sum $\sum_{\mathbf{n}}P^{(B)}(\mathbf{n})=1$ into the integral  $I_M$ of $\rho $ in Eq. (\ref{E11}) over $\xi_1,\ldots,\xi_r$ with  $0\le \xi_i\le \sqrt{M}$. The latter is exponentially close to $1$ for  $M\gg 1$: $I_M = 1-e^{-\mathcal{O}(M)}$.)

%%%%%%%%%%%%%%%%%%%%
\section{Proof of the Theorems }
\label{sec3}
%%%%%%%%%%%%%%%%%
\subsection{The binary classical case}
\label{sec3A}
Let us first consider the classical binary case $r=2$. Denote by $K_2$  the Kullback-Leibler  divergence 
\begin{equation}
K_2(x|q) = x \ln\left(\frac{ x}{q}\right) +(1-x) \ln \left(\frac{ 1-x}{1-q}\right).
\label{E12}\end{equation}
Using   Stirling's formula $ n! = \sqrt{2\pi (n+\theta_n)}(n/e)^n$, where  $\frac{1}{6}<\theta_n<1.77$ for $ n\ge 1$  and $\theta_0 = \frac{1}{2\pi}$
 \cite{Mortici}, we have   
\begin{eqnarray}
\fl \qquad P^{(D)}_{N,M}(n|K) &= &\left[\frac{1+ \theta_N/N}{2\pi N (x +\theta_n/N) (1-x + \theta_{N-n}/N)}\right]^\frac12\left(\frac{x}{q}\right)^{-n}
\left(\frac{1-x}{1-q}\right)^{-N+n}
\nonumber\\
\fl & = &  \frac{\exp\left\{ -NK_2(x|q) \right\}}{\sqrt{2\pi Nq(1-q)}}\left[1+\mathcal{O}\left(\frac{1}{N^{\frac13+\epsilon}}\right)\right],
 \label{E13}\end{eqnarray}
since from Eq. (\ref{E6})  
\[
\frac{x}{q}\frac{1-x}{1-q} \ge \left(1 - \frac{A}{qN^{\frac13+\epsilon}}\right)\left(1 - \frac{A}{(1-q)N^{\frac13+\epsilon}}\right)  = 
1 + \mathcal{O}\left(\frac{1}{N^{\frac13+\epsilon}}\right).
\]
 By  expanding the Kullback-Leibler divergence (\ref{E12}) using  Eq. (\ref{E6}),
\begin{equation}
K_2(x|q) = \frac{(x-q)^2}{2q(1-q)} +\mathcal{O}\left(\frac{1}{N^{3\epsilon}}\right),
\label{E14}\end{equation}
and substituting  the result in Eq. (\ref{E13}) we get  Eq. (\ref{E7}) for the binary classical case.  

To show Eq.~(\ref{E8}) consider the first line of Eq. (\ref{E13}), valid for all $0\le x \le 1$, and  observe that $(x+\theta_{n}/N)(1-x+\theta_{N-n}/N)\ge (2\pi N)^{-2}$. We  obtain
\begin{equation}
P^{(D)}_{N,M}(n|K) \le 2\pi \sqrt{N} \exp\left\{ -N K_2(x|q) \right\}\left[1+\mathcal{O}\left(\frac{1}{N}\right)\right].
\label{P2D}\end{equation}
Then using  Pinsker's inequality \cite{Pinsker}
\begin{equation}
K_2(x|q) \ge (x-q)^2
\label{E15}\end{equation}
and  that by Eq. (\ref{E6}) $|x-q| > A N^{-\frac13-\epsilon}$ for $\epsilon <1/6$ we   obtain the required  scaling of Eq. (\ref{E8}) from  Eq.~(\ref{P2D}):
\begin{equation}
\fl P^{(D)}_{N,M}(n|K) \le 2\pi \sqrt{N} \exp\left\{ -A^2N^{\frac13-2\epsilon} \right\}\left[1+\mathcal{O}\left(\frac{1}{N}\right)\right] 
 = \mathcal{O}\left(\sqrt{N} \exp\left\{-A^2 N^{\frac13-2\epsilon}\right\}\right).
\label{E16}\end{equation}

%%%%%%%%%%%%%%%%%%%%%%
\subsection{The $r$-bin classical case}
\label{sec3B}
Let us consider the average probability for the  general $r$-bin classical case.  We can employ the factorization  into $r-1$ binary probabilities given by  Eqs. (\ref{E3})-(\ref{E4}). First of all, let us show the equivalence of Eq. (\ref{E6}) to the following set of conditions (see Eq. (\ref{E5})):
\begin{equation}
|n_l - N_l \bar{q}_l| \le \bar{A} N^{\frac23-\epsilon}, \quad \bar{A}>0, \quad  l = 1,\ldots, r-1. 
\label{E17}\end{equation}
To this end  it is enough to observe that  
\begin{equation}
n_l-N_l \bar{q}_l = n_l - Nq_l + \bar{q}_l \sum_{i=1}^{l-1} (n_i - N q_i), \quad l = 1,\ldots,r-1.
\label{E18}\end{equation}
Indeed, the relations in Eq. (\ref{E18}) are invertible, whereas $n_r - Nq_r = - \sum_{l=1}^{r-1}(n_l - Nq_l)$. To  prove  the classical $r$-bin case in theorems 1 and 2 one  can proceed  as follows.  If  Eq. (\ref{E6}) is satisfied for all $i$, then so is Eq. (\ref{E17}). Using  Eq. (\ref{E13}) for  the binary case into Eq. (\ref{E3})  we have 
\begin{equation}
\!\!\!\!\!\!\!\!\!\!\!\!\!\!\!\!\!\! P^{(D)}(\mathbf{n}|\mathbf{K}) = \prod_{l=1}^{r-1} P^{(D)}_{N_l,M_l}(n_l|K_l) =\prod_{l=1}^{r-1} \frac{\exp\left\{ -N_lK_2(\bar{x}_l|\bar{q}_l) \right\}}{\sqrt{2\pi N_l\bar{q}_l(1-\bar{q}_l)}}\left[1+\mathcal{O}\left(\frac{1}{N^{\frac13+\epsilon}}\right)\right]
\label{E19}\end{equation}
From Eq. (\ref{E5})  we get
\begin{equation}
N_l\bar{q}_l(1-\bar{q}_l) = Nq_l \frac{1-\sum_{i=1}^lq_i}{1-\sum_{i=1}^{l-1}q_i}\left(1 - \frac{\sum_{i=1}^l n_i/N - q_i}{1-\sum_{i=1}^{l-1}q_i} \right),
\label{E20}\end{equation}
therefore the denominator in Eq. (\ref{E19}) becomes equal to that in Eq. (\ref{E7}) to the necessary error, i.e.,
\begin{equation}
\prod_{l=1}^{r-1} N_l\bar{q}_l(1-\bar{q}_l) = N^{r-1}\prod_{i=1}^r q_i \left[1 + \mathcal{O}\left(\frac{1}{N^{\frac13+\epsilon}}\right) \right]
\label{E21}\end{equation}
(since $1/3 +\epsilon > 3\epsilon$ for $\epsilon < 1/6$, the error conforms with that in Eq. (\ref{E7})).  In its turn, the  term in the exponent in Eq. (\ref{E19})  can be reshaped using the following identity for the Kullback-Leibler divergence (see  \ref{appB})
\begin{equation}
\sum_{l=1}^{r-1} N_l K_2(\bar{x}_l|\bar{q}_l) = N\sum_{i=1}^r x_i \ln \left(\frac{ x_i}{q_i}\right)\equiv N K_r(\mathbf{x}|\mathbf{q}). 
\label{E22}\end{equation}
 By expanding both sides of Eq. (\ref{E22}) into powers of $x_i-q_i$  and comparing the terms to the leading  order in $N$ we obtain  the following asymptotic identity 
\begin{equation}
\sum_{l=1}^{r-1} N_l \frac{(\bar{x}_l-\bar{q}_l)^2}{2\bar{q}_l(1-\bar{q}_l)} = N \sum_{l=1}^{r-1} \frac{(x_l-q_l)^2}{2q_l} +  \mathcal{O}\left(\frac{1}{N^{3\epsilon}}\right).
\label{E23}\end{equation}
 Substituting  Eqs. (\ref{E21}) and (\ref{E23}) into  Eq. (\ref{E19}) we obtain  Eq. (\ref{E7}) for the $r$-bin  classical case.  

 To show  (\ref{E8}) for the $r$-bin classical case,  let us select  the factorization (\ref{E3}) such that $i=1$ is the  first violation of  Eq.~(\ref{E6}).  Then the probability $P^{(D)}_{N,M}(n_1|K_1)$  appearing in Eq.~(\ref{E3}) satisfies Eq.~ (\ref{E8}),  as proven  above in the binary case. This observation results in Eq. (\ref{E8}) for the $r$-bin classical case and concludes the proof of the  theorems in the   classical case. 

One important note. In the course of the proof of the theorems for the $r$-bin case we have also shown that the asymptotic Gaussian for the $r$-bin case is a product of the asymptotic Gaussians for the $r-1$ binary cases to the same accuracy as in Eq. (\ref{E7}), i.e., due to the equivalence of Eqs. (\ref{E6}) and (\ref{E17})  the general case follows from the binary case. 

%%%%%%%%%%%%%%%%%%%%%%
\subsection{The quantum  case}
\label{sec3C}
 Let us consider the quantum factor $Q^{(\pm)} (\mathbf{n}|\mathbf{K})$ (recall that $+$ is for bosons and $-$ is for fermions) introduced in the second line in Eq. (\ref{E2}), which accounts for the correlations between the indistinguishable  particles due to their quantum statistics. By the following asymptotic identity  \cite{GL}
 \begin{equation}
\prod_{l=0}^n \left[1\pm \frac{l}{m}\right] = \left( 1\pm \frac{n}{m}\right)^{n\pm m +1/2} e^{-n}\left[1+\mathcal{O}\left(\frac{n}{m(m\pm n)}\right)\right]
\label{E24}\end{equation}
(see  also \ref{appC}) when $N,M\to \infty$ we get  for $n_i$ satisfying Eq. (\ref{E6}) (i.e., $n_i = N\left[q_i + \mathcal{O}(N^{-\frac13-\epsilon})\right]\to \infty$) 
 \begin{eqnarray}
\fl && Q^{(\pm)}(\mathbf{n}|\mathbf{K})  \equiv  \frac{\prod_{i=1}^r\left(\prod_{l=0}^{n_i-1}\left[1\pm l/K_i\right]\right)}{\prod_{l=0}^{N-1}\left[1\pm l/M\right]} = \frac{\prod_{i=1}^r\left(\prod_{l=0}^{n_i}\left[1\pm l/K_i\right]\right)}{\prod_{l=0}^{N}\left[1\pm l/M\right]} \frac{1\pm \frac{N}{M}}{\prod_{i=1}^{r}\left[1\pm n_i/K_i\right]} \nonumber\\
\fl && = \frac{\prod_{i=1}^r\left(1\pm n_i/K_i\right)^{n_i\pm K_i -1/2}}{\left(1\pm N/M\right)^{N\pm M -1/2}} \left[1 + \mathcal{O}\left( \sum_{i=1}^r\frac{n_i}{K_i(K_i\pm n_i)}+\frac{N}{M(M\pm N)}\right)\right]
\label{E25}\end{eqnarray}
(with the upper signs for bosons and the lower ones  for fermions). Note that in the case of fermions $n_i \le K_i$  (for $n_i>K$  quantum factor is equal to zero). Now, let us clarify the order of the error in Eq. (\ref{E25}). From Eq. (\ref{E6}) we get
\[
K_i \pm n_i = (M\pm N)q_i \mp[ Nq_i - n_i ]  \ge \left|\frac{1\pm \alpha}{\alpha} q_iN - A N^{2/3-\epsilon}\right|.
\] 
Thus  we can estimate
\[
\frac{n_i}{K_i(K_i\pm n_i)} = \mathcal{O}\left( \frac{\alpha^2}{(1\pm \alpha)N}\right).
\]
Taking this into account, let us rewrite   Eq. (\ref{E25}) as follows 
\begin{equation}
Q^{(\pm)}(\mathbf{n}|\mathbf{K}) =  \frac{\exp\{(N\pm M)K_r(\mathbf{X^{(\pm)}}|\mathbf{q}) \}}{(1\pm \alpha)^{\frac{r-1}{2}} \prod_{i=1}^r \sqrt{\frac{X^{(\pm)}_i}{q_i}}} \left[1+ \mathcal{O}\left(\frac{\alpha^2}{(1\pm \alpha)N}\right)\right],
\label{E26}\end{equation}
where $K_r$ is defined in Eq. (\ref{E22}) and we have introduced new variables $X^{(\pm)}_i$ (analogs of $x_i$ of Eq. (\ref{E5}) in the quantum case)
\begin{equation}
\fl \quad X^{(\pm)}_i \equiv \frac{K_i \pm n_i}{ M \pm  N} = \frac{q_i \pm \alpha x_i}{1\pm \alpha}, \quad 0\le X^{(\pm)}_i \le 1,\quad X^{(\pm)}_i  - q_i = \frac{\pm\alpha}{1\pm \alpha}(x_i - q_i). 
\label{E27}\end{equation}
Now, if Eq. (\ref{E6}) is satisfied,  we can separate the leading order in the quantum factor by expanding the Kullback-Leibler divergence (similar as in  Eq. (\ref{E23})), whereas  in the denominator in Eq. (\ref{E26})  we have   \mbox{$X^{(\pm)}_i/q_i = 1 +\mathcal{O}\left(\frac{\alpha N^{-1/3-\epsilon}}{1\pm\alpha}\right)$}. By selecting  the leading order error for $0<\epsilon <1/6$   we get (recall that $\sigma = +$ for bosons and $\sigma = -$ for fermions):
\begin{equation}
\fl Q^{(\sigma)}(\mathbf{n}|\mathbf{K}) = \frac{\exp\left\{N \frac{\sigma \alpha}{1+\sigma \alpha}\sum_{i=1}^r\frac{(x_i-q_i)^2}{2q_i}\right\}}{(1+\sigma \alpha)^{\frac{r-1}{2}} } \left[1+ \mathcal{O}\left(\frac{\alpha\delta_{\sigma,+}}{N}+\frac{\alpha^3}{(1+\sigma \alpha)^3N^{3\epsilon}}\right)\right].
\label{E29}\end{equation}
For bosons the first term in the error on the right hand side of Eq. (\ref{E29})  can dominate the second in the high-density case $\alpha\to \infty$, thus we have to keep it. To obtain  Eq. (\ref{E7}) in the quantum case one can just  multiply the  result of Eq. (\ref{E7}) for the classical probability, proven in section \ref{sec3B}, by the quantum factor in 
Eq.~(\ref{E29}) and select  the leading order error terms (observing the possibility that $\alpha\to\infty$ for bosons and $\alpha\to 1$ for fermions). This proves theorem 1 in the quantum case. 

One important observation is in order. The $r$-bin   quantum  factor $Q^{(B,F)}$ of Eq. (\ref{E25}) is simply a product of the $r-1$  binary  quantum factors $Q^{(B,F)}_{N_l,M_l}$ defined similar as in Eq. (\ref{E25}),  but with $M_l$ and  $N_l$ as in the factorization formula (\ref{E3}) and $\bar{X}^{(\pm)}_l$ defined as in Eq. (\ref{E5}).  This fact simply follows from the factorization formula (\ref{E3}) valid in the classical and quantum cases. The same factorization  is valid also for the leading order of the respective quantum factors, up  to the error term in   Eq. (\ref{E29}).  In fact, one proceed to prove Eq. (\ref{E7}) in the general $r$-bin case using the binary case, similar as it was done in section \ref{sec3B}.  Indeed, there is  the following  identity for the Kullback-Leibler divergence (an analog of Eq.  (\ref{E22}))
\begin{equation}
\fl \sum_{l=1}^{r-1} (N_l \pm M_l) K_2(\bar{X}^{(\pm)}_l|\bar{q}_l) = (N\pm M) \sum_{l=1}^r X^{(\pm)}_l \ln\left(\frac{ X^{(\pm)}_i}{q_i}\right)=(N\pm M)K_r(\mathbf{X}^{(\pm)}|\mathbf{q}),
\label{E30}\end{equation}
which is proved via  the same steps  as the respective identity (\ref{E22})  in the classical case (\ref{appB}). 

Let us now prove theorem 2 in the quantum case.  The quantum result in Eq. (\ref{E8}) can be  shown by reduction to the binary case, similar as in section \ref{sec3B}, where $i=1$ is the first  index of violation of Eq. (\ref{E6}).   Consider  the respective quantum probability $P^{(B,F)}_{N,M}(n_1|K_1)=P^{(D)}_{N,M}(n_1|K_1) Q^{(B,F)}_{N,M}(n_1|K_1) $ which enters the factorization (\ref{E3}) in the quantum case (below we drop the subscript $1$  for simplicity).  

Let us first focus on  the case of bosons.  From Eq. (\ref{C10}) of \ref{appC}  for $M\gg 1$ (see Eqs. (\ref{E12}) and (\ref{E27}))  we get
\begin{equation}
\fl \qquad Q^{(B)}_{N,M}(n|K) < (1+\alpha)\exp\left\{ N\left(1+\frac{1}{\alpha}\right)K_2(X^{(+)}|q)\right\}\left[ 1+\mathcal{O}\left(\frac{1}{M}\right)\right].
\label{E31}\end{equation}
The quantum probability $P^{(B)}_{N,M}(n|K) = P^{(D)}_{N,M}(N|K)Q^{(B)}_{N,M}(n|K)$ involves a  combination of two Kullback-Leibler divergencies, see Eqs. (\ref{P2D}) and (\ref{E31}),
\begin{equation}
\fl \qquad NK_2(x|q) - N\left(1+\frac{1}{\alpha} \right)K_2(X^{(+)}|q) = N\left[K_2(x|X^{(+)}) +\frac{1}{\alpha}  K_2(q|X^{(+)})\right]. 
\label{E32}\end{equation}
By using Pinsker's inequality (\ref{E15})  and Eq. (\ref{E27}) we obtain 
\begin{equation}
 \fl \qquad  K_2(x|X^{(+)}) +\frac{1}{\alpha}  K_2(q|X^{(+)}) > \left( X^{(+)}-x\right)^2 + \frac{1}{\alpha}\left(X^{(+)}- q\right)^2 = \frac{(x-q)^2}{1+\alpha}.
\label{E33}\end{equation}
Finally, taking into account that by our assumption $|x-q| > A N^{-1/3-\epsilon}\,$ and that $\alpha $ is fixed  in theorem 2,  from Eqs. (\ref{P2D}), (\ref{E31})-(\ref{E34}) we get the required  estimate 
\begin{equation}
P^{(B)}_{N,M}(n|K) < 2\pi \sqrt{N}(1+\alpha)\exp\left\{-\frac{{A}^2}{1+\alpha}N^{\frac13 -2\epsilon} \right\}\left[ 1+\mathcal{O}\left(\frac{1}{N}\right)\right].
\label{E34}\end{equation}

Now let us turn to the case of fermions. First of all, we have to consider the maximal possible  count number  $n=K$   (or $N-n = M-K$ which  amounts to renaming the variables, but not both since $N<M$).  In this case, under the condition that $\alpha$ is fixed, the average quantum counting probability reads (see Eq. (\ref{E2}))
\begin{eqnarray}
\fl P^{(F)}(n|K) &=& \frac{N!}{(M-N+1)\ldots M} \frac{(M-K-[N-K]+1)\ldots (M-K)}{(N-K)!} \nonumber\\
\fl &=&\left( \frac{N}{M}\right)^K \prod_{l=1}^{K-1} \frac{1-l/K}{1-l/M} < \left( \frac{N}{M}\right)^K =   \exp\left\{-N \frac{\ln(1/\alpha)}{q}\right\}
\label{E35}\end{eqnarray}
i.e., falls faster with $N$  than the estimate in  Eq. (\ref{E8}). Consider now  the opposite case $n\le K-1$ and $N-N\le M-K-1$. We have in this case $X^{(-)}, 1-X^{(-)}  \ge 1/(M-N) = \frac{\alpha}{(1-\alpha)N}$. Therefore,    from Eq. (\ref{C10}) of \ref{appC} we get
\begin{eqnarray}
\fl Q^{(F)}_{N,M}(n|K)& <& \frac{q(1-q)\exp\{-N(1/\alpha-1)K_2(X^{(-)}|q)\}}{(1-\alpha)^2X^{(-)}(1-X^{(-)})}\left[ 1+\mathcal{O}\left(\frac{1}{M} \right)\right]\nonumber\\
\fl &=& \mathcal{O}\left(N^2  \exp\left\{-\frac{{A}^2}{1-\alpha} N^{\frac13-2\epsilon}\right\}\right),
\label{E36}\end{eqnarray}
where we have expanded the Kullback-Leibler divergence as in Eq. (\ref{E14}),  used  Eq. (\ref{E27})  and Pinsker's inequality (\ref{E15}) together with the assumption $|x-q| > {A} N^{-1/3-\epsilon}\,$ for $\epsilon <1/6$. Recalling the respective classical bound (\ref{E16})  we get Eq. (\ref{E8}) for fermions. This concludes the proof of theorem 2 in the quantum case. 

Finally, as in the classical case, in the quantum case the asymptotic Gaussian for the  $r$-bin partition in Eq. (\ref{E7})  is a product of the asymptotic Gaussians for the binary partitions, which appear in Eq. (\ref{E3}), to the accuracy of the approximation in Eq. (\ref{E7}), due to the analogous identity Eq. (\ref{E30}). This fact relates the statements of   theorems 1 and 2 to those of the binary case via the equivalence of Eqs. (\ref{E6}) and (\ref{E17}).

\section{Abitrary (mixed)  input state}
\label{sec4}

In section \ref{sec2} in the formulation of theorems 1 and 2  we have assumed  a Fock input state $|\bn,in\rangle = |n_1,\ldots,n_M;in\rangle $ of $N$ indistinguishable  identical particles (where for bosons  $n_k$ is arbitrary, whereas for fermions $n_k\le 1$). However, it is easy to see that the theorems generalize  to an arbitrary input state
\begin{equation}
\rho = \sum_{\bn,\blm} \rho_{\bn,\blm} |\bn,in\rangle \langle \blm,in|,
\label{AE1}\end{equation}
where the summation is over $|\bn| = |\blm| = N$ ($|\bn| \equiv n_1+\ldots n_M$). Let us consider bosons first. Using the expansion of the input Fock state $|\bn,in\rangle $ over the output $|\bls,out\rangle$ \cite{AA}
\begin{equation}
|\bn,in\rangle = \sum_{\bls}\frac{1}{\sqrt{\bn!\bls!}}\mathrm{per}(U[\bn|\bls])|\bls,out\rangle,
\label{AE2}\end{equation}
where the summation is over all $|\bls|=N$, $\bn! \equiv n_1!\ldots n_M!$, and per($\ldots$) denotes the matrix permanent \cite{Minc}, in our case  of the submatrix of the $M$-port matrix $U$ built on the rows and columns corresponding to the occupations $\bn$ and $\bls$, respectively.   Given the input state in Eq. (\ref{AE1}), the average probability to detect an output configuration $\bl$, corresponding to occupations $\bls$, reads
\begin{equation}
p^{(B)}(\bl|\rho)  =  \frac{1}{\bls!} \sum_{\bn,\blm}  \frac{\rho_{\bn,\blm}}{\sqrt{\bn!\blm!}}\langle \mathrm{per}(U[\bn|\bls]) \left(\mathrm{per}(U[\blm|\bls])\right)^*\rangle.
\label{AE3}\end{equation}
Let us evaluate the average by expanding the matrix permanents 
\begin{eqnarray}
\label{AE4}
& & \langle \mathrm{per}(U[\bn|\bls]) \left(\mathrm{per}(U[\blm|\bls])\right)^*\rangle =  \langle\sum_{\sigma_{1,2}\in \mathcal{S}_N} \prod_{i=1}^N U_{k^{}_{\sigma_1(i)},l^{}_i}U^*_{k^\prime_{\sigma_2(i)},l^{}_i} \rangle\nonumber\\
& & = \sum_{\sigma_{1,2}\in \mathcal{S}_N} \sum_{\nu,\tau\in \mathcal{S}_N}\mathcal{W}(\tau\nu)\prod_{i=1}^N\delta_{k^\prime_{\sigma_2(i)},k^{}_{\sigma_1\nu(i)}}\delta_{l_i,l_{\tau(i)}}\nonumber\\
& =& \delta_{\bn,\blm} \sum_{\sigma_{1,2}\in \mathcal{S}_N} \sum_{\nu,\tau\in \mathcal{S}_N} \sum_{\chi\in\mathcal{S}_\bn} \sum_{\mu\in \mathcal{S}_\bls} \mathcal{W}(\tau\nu)\delta_{\sigma_1\nu\sigma^{-1}_2,\chi} \delta_{\tau,\mu}\nonumber\\
& & = \delta_{\bn,\blm} \sum_{\sigma_{1,2}\in \mathcal{S}_N} \sum_{\mu\in \mathcal{S}_\bls}\sum_{\chi\in\mathcal{S}_\bn} \mathcal{W}(\mu \sigma^{-1}_1\chi \sigma_2) = \delta_{\bn,\blm} \bls!\bn! N! \sum_{\sigma\in \mathcal{S}_N} \mathcal{W}(\sigma)\nonumber\\
&& =   \frac{\delta_{\bn,\blm} \bls!\bn!N!}{(M+N-1)\ldots M},
\end{eqnarray}
where $\bk=(k_1,\ldots,k_N)$ and $\bk^\prime=(k^\prime_1,\ldots,k^\prime_N)$ are the input ports corresponding to the occupations $\bn$ and $\blm$, respectively, $\bl = (l_1,\ldots,l_N)$ are the output ports corresponding to the occupations $\bls$,  $\mathcal{S}_N$ is the group of permutations of $N$ elements (the symmetric group), whereas $\mathcal{S}_\bn \equiv \mathcal{S}_{n_1}\otimes \ldots \otimes \mathcal{S}_{n_M}$,  $\mathcal{W}$ is the Weingarten function of the unitary group \cite{W1,W2}, and  $\delta_{\bn,\blm} \equiv \prod_{i=1}^M \delta_{n_i,m_i}$.  We have used the known expression for the last sum on the right hand side of Eq. (\ref{AE4}) (derived in  the Supplemental material to Ref. \cite{BB}). 

Eq. (\ref{AE4}) tells us that the non-diagonal elements of the mixed state in Eq. (\ref{AE1}) do not contribute to the average probability, if the averaging is performed over the Haar-random unitary matrix $U$ (the ratio on the right hand side of Eq. (\ref{AE4}) is the average probability $p^{(B)}(\bl|\bk)$, see section \ref{sec2}). Since, theorems 1 and 2 hold  for any Fock input state $|\bn,in\rangle$, we conclude that they hold for the general input of Eq. (\ref{AE1}).  

For fermions, an analog of  Eqs. (\ref{AE3}) and (\ref{AE4})  (in this case $m_i,n_i,s_i\le 1$) are obtained by replacing the permanent by the determinant, which results in  the appearance of the  sign functions $\mathrm{sgn}(\sigma_{1,2})$ and $\mathrm{sgn}(\sigma)$  ($\sigma = \sigma_1\sigma_2$) in Eq. (\ref{AE4}) where there are $\sigma_{1,2}$ and $\sigma$. In this case the last summation in Eq. (\ref{AE4}) reads  N!$\sum_{\sigma\in \mathcal{S}_N} \mathrm{sgn}(\sigma)\mathcal{W}(\sigma)   =  \frac{N!}{M \ldots (M-N +1)}$  (see  the Supplemental material to Ref. \cite{BB}) i.e.,  we get the average probability $p^{(F)}(\bl|\bk)$.  The same conclusion holds.

\section{Conclusion}
\label{sec5}

We have given a rigorous  formulation of the results  on the asymptotic form of the  average  counting probability  of identical particles in the binned-together output ports of the Haar-random multiports, presented recently  in Ref. \cite{GL} with only a heuristic derivation and some  numerical evidence. The key observation was that, both in the classical and quantum cases, there is a convenient factorization of the average probability for the $r$-bin case into $r-1$   average counting probabilities for the two-bin case. Moreover, the results of Ref. \cite{GL} were extended to an arbitrary mixed input state of $N$ indistinguishable particles.

In the classical case, we have shown that the de Moivre-Laplace theorem, which provides an asymptotic form of the binary average counting  probability, actually applies also to the $r$-bin case via the above factorization. The asymptotic Gaussian form also satisfies the mentioned factorization to an error of the same   order as in the   Moivre-Laplace theorem. Finally, though we have considered a physical model involving a random unitary  multiport, where the probabilities of $r$ events are rationals (each probability equal to a fraction of  the respective number of ports),   the results  apply for a general multinomial distribution with arbitrary such probabilities (since the factorization is derived for the general probabilities).  

Our  primary interest, however, was the quantum case,  when there are correlations between the identical particles  due to their quantum statistics. We have formulated and proven a quantum analog of the de Moivre-Laplace theorem for the indistinguishable  identical bosons and fermions  (and  generalized it to the $r$-bin case), where again the binary case applies to the $r$-bin case by the above mentioned factorization (and, similarly to the classical case, the asymptotic Gaussian also satisfies the same factorization to the order of the approximation error).    Therefore, besides giving a rigorous formulation of the recently discovered quantum asymptotic Gaussian law, we have also provided an illuminating insight  on how the general $r$-bin case reduces to the binary case.

Our results have immediate applications for the counting probability (in the binned-together output modes) of   identical particles propagating in  the  disordered media,  chaotic cavities, and  also for the  scattershot version of the Boson Sampling.

\section{Acknowledgements}
The research was supported by the National Council for Scientific and Technological Development (CNPq) of Brazil,  grant  304129/2015-1, and by  the S{\~a}o Paulo Research Foundation   (FAPESP), grant 2015/23296-8.

\appendix 

 %%%%%%%%%%%%%%%%%%%%%%%%%%%%%%
\section{The factorization of the counting probability}
\label{appA}

Consider first the classical case. We have (for general $q_1, \ldots, q_r$)
\begin{eqnarray}
\fl \quad && P(n_1,\ldots,n_r|q_1,\ldots,q_r) \equiv \frac{N!}{\prod_{i=1}^r n_i!} \prod_{i=1}^r q_i^{n_i} = \frac{N!}{n_1! (N-n_1)!} q_1^{n_1}(1-q_1)^{N-n_1}\nonumber\\
\fl && \times \frac{(N-n_1)!}{\prod_{i=2}^r n_i!} \prod_{i=2}^r \left(\frac{q_i}{1-q_1}\right)^{n_i} =\ldots \nonumber\\
\fl && =  P(n_1,N-n_1|q_1,1-q_1)\ldots P(n_{r-1},N_{r-1}-n_{r-1}|\bar{q}_{r-1},1-\bar{q}_{r-1}),
\label{A1}\end{eqnarray}
here the dots denote the  sequential factorization (similar to that in the first line),  where  have taken into account the definitions in Eqs. (\ref{E4}) and (\ref{E5}) and that 
\begin{eqnarray*}
&& \frac{q_2}{1-q_1} = \bar{q}_2,\quad \frac{q_3}{(1-q_1)(1-\bar{q}_2)} = \frac{q_3}{1-q_1-q_2} = \bar{q}_3, \\
&&  \frac{q_4}{(1-q_1)(1-\bar{q}_2)(1-\bar{q}_3)} = \frac{q_4}{1-q_1-q_2 -q_3 } = \bar{q}_4, \ldots.
\end{eqnarray*}
Eq. (\ref{A1}) implies the stated factorization for the classical probability $P^{(D)}(\mathbf{n}|\mathbf{K})$. 

Now let us consider the quantum case. We have 
\begin{eqnarray}
\fl  \quad & & P^{(B,F)}(n_1,\ldots,n_r|K_1,\ldots,K_r) = \frac{N!}{\prod_{i=1}^r n_i!} \frac{(M-1)!}{(M\pm N \mp 1)!}\prod_{i=1}^r\frac{(K_i\pm n_i \mp 1)!}{(K_i-1)! }
\nonumber\\
\fl && = \frac{N!}{n_1! (N-n_1)!} \frac{(M-1)!}{(M\pm N\mp1)!}\frac{(K_1\pm n_1 \mp1)!}{(K_1-1)!}\frac{(M-K_1\pm [N-n_1]\mp 1)!}{(M-K_1-1)!} \nonumber\\
\fl && \times P^{(B,F)}(n_2,\ldots,n_r|K_2,\ldots,K_r) = \ldots \nonumber\\
\fl && = P^{(B,F)}(n_1,N-n_1|K_1,M-K_1)\ldots P^{(B,F)}(n_{r-1},N_{r-1}-n_{r-1}|K_{r-1},M_{r-1}-K_{r-1}),\nonumber\\
\fl \label{A2}\end{eqnarray}
where again the sequential factorization was employed with  the definitions in Eq. (\ref{E4}) (for instance, in the second factorization we have $\sum_{i=2}^rK_i = M - K_1= M_2$ and $\sum_{i=2}^r n_i =  N- n_1= N_2 $).

%%%%%%%%%%%%%%%%%%%%%%%%%%%%%%%%
\section{An identity for  the Kullback-Leibler divergence }
\label{appB}

Let us rewrite the Kullback-Leibler divergence in Eq. (\ref{E19})  (see also the definitions in Eq. (\ref{E5})) as follows 
\begin{equation}
N_lK_2(\bar{x}_l|\bar{q}_l) = N\left\{ x_l \ln\left(\frac{x_l}{q_l}\right) + Z_l \ln\left(\frac{Z_l}{Q_l}\right) - Z_{l-1} \ln\left(\frac{Z_{l-1}}{Q_{l-1}}\right)   \right\},
\label{B1}\end{equation}
where we have denoted 
\begin{equation}
Z_l \equiv 1-\sum_{i=1}^{l}x_i,\quad Q_l \equiv 1-\sum_{i=1}^{l}q_i.
\label{B2}\end{equation}
 Now it is easy to see that due to the form of the last two terms in Eq. (\ref{B1})   the sum of the Kullback-Leibler divergencies as in Eq. (\ref{B1}) with $l=1,\ldots,r-1$ which appear in Eq. (\ref{E19}) give
 \begin{equation}
\fl \sum_{l=1}^{r-1}  N_lK_2(\bar{x}_l|\bar{q}_l) = N\left\{  \sum_{l=1}^{r-1} x_l  \ln\left(\frac{x_l}{q_l}\right) + Z_{r-1}\ln\left(\frac{Z_{r-1}}{Q_{r-1}}\right)\right\} = N  \sum_{l=1}^{r} x_l\ln\left(\frac{x_l}{q_l}\right), 
\label{B3}\end{equation}
since $Z_{r-1} = x_r$ and $Q_{r-1} = q_r$.

%%%%%%%%%%%%%%%%%%%%%%%%%%%%%%%%%%%%%
\section{Asymptotic form of  $\prod_{l=1}^n\left[1\pm \frac{l}{m}\right]$}
\label{appC}
We will use the second-order Euler's summation formula \cite{EulerSum}:
\begin{equation}
\!\!\!\!\!\!\!\!\!\!\!\!\!\!\!\!\!\!\!\!\!\!\!\! \sum_{l=1}^n f(l)  = \int\limits_1^ndx\, f(x) + \frac{f(n)+f(1)}{2} + \frac{f^{(1)}(n) - f^{(1)}(1)}{12} - \frac{1}{2} \int\limits_1^n dx\, P_2(x) f^{(2)}(x),\quad
\label{C1}\end{equation}
where $|P_2(x)| \le 1/6 $.  Setting $f(x) = \ln(1\pm x/m)$ (in our case $1\le x \le n$), we observe that
\begin{equation}
f^{(1)}(n) - f^{(1)}(1) = \frac{1}{n\pm m} - \frac{1}{1\pm m} = \mathcal{O}\left(\frac{n}{m(m\pm n)}\right)
\label{C2}\end{equation}
and 
\begin{equation}
\left|\int\limits_1^n dx\, P_2(x) f^{(2)}(x)\right| \le \frac{|f^{(1)}(n) - f^{(1)}(1)|}{6}.
\label{C3}\end{equation}
Then by integrating 
\begin{equation}
\fl \quad \int\limits_1^n dx\,\ln \left(1\pm \frac{l}{m}\right) = (n\pm m)\ln \left(1\pm \frac{n}{m}\right) - (1\pm m) \ln \left(1\pm \frac{1}{m}\right) - n+1
\label{C4}\end{equation}
 and  using  Eqs. (\ref{C1})-(\ref{C3})   we obtain Eq. (\ref{E24}).  

One can also find the upper and lower bounds using that the two involved in the derivation of Eq. (\ref{E24})  functions $f^{(\pm)} = \ln \left(1\pm x/m\right)$ are strictly monotonous. Let us first consider $f^{(+)}(x)$. By the  geometric consideration similar to  that of Ref.  \cite{EulerSum} one can easily establish that
\begin{equation}
\int\limits_1^n dx\,f^{(+)}(x) < \sum_{l=1}^n f^{(+)}(l) < \int\limits_1^n dx\,f^{(+)}(x) + f^{(+)}(n). 
\label{C5}\end{equation}
Taking $\hat{f}^{(+)}(x) \equiv  -f^{(-)}(x)$ and using Eq. (\ref{C2}) we get
\begin{equation}
\int\limits_1^n dx\,f^{(-)}(x) +  f^{(-)}(n)< \sum_{l=1}^n f^{(-)}(l) < \int\limits_1^n dx\,f^{(-)}(x). 
\label{C6}\end{equation}
Eqs. (\ref{C4})-(\ref{C6}) allow to get the announced bounds. First of all, for $m\gg 1$ (in our case $K_i$ or $M$  take place of $m$) we can approximate
\begin{equation}
\fl \quad  \left(1\pm \frac{1}{m}\right)^{1\pm m} = \exp\left\{ (1\pm m) \sum_{p=1}^\infty \frac{(-1)^{p-1}}{p} \frac{(\pm1)^p}{m^p}\right\} =  e\left[ 1+\mathcal{O}\left(\frac{1}{m}\right)\right].
\label{C7}\end{equation}
Therefore, from Eqs.  (\ref{C4})-(\ref{C7}) we obtain (the upper  digit in the parenthesis is for the plus   sign,  while the lower choice is  for the minus sign):
\begin{equation}
\prod_{l=1}^n \left(1\pm \frac{l}{m}\right) <\left(1\pm \frac{n}{m}\right)^{n\pm m + \{{1 \atop 0} \}}e^{-n}\left[ 1+\mathcal{O}\left(\frac{1}{m}\right)\right],
\label{C8}\end{equation}
\begin{equation}
\prod_{l=1}^n \left(1\pm \frac{l}{m}\right) \ge \left(1\pm \frac{n}{m}\right)^{n\pm m + \{{0 \atop 1} \}}e^{-n}\left[ 1+\mathcal{O}\left(\frac{1}{m}\right)\right].
\label{C9}\end{equation}

Let us now  find the  bounds on the quantum factor in Eq. (\ref{E25}).  Using the definition of the Kullback-Leibler divergence (\ref{E22})  and the quantities $X_i^{(\pm)} $ defined in Eq. (\ref{E27})  for $M\gg 1$  we obtain from Eqs. (\ref{C8}) and (\ref{C9})  (with $\sigma=+$ for bosons, the upper line in the parenthesis, and $\sigma=-$ for fermions, the lower line in the parenthesis):
\begin{equation}
Q^{(\sigma)} < Q^{(\sigma)}_{as} \left\{ {1+\alpha \atop (1-\alpha)^{-r} \prod_{i=1}^r\left( \frac{X_i^{(-)}}{q_i}\right)^{-1} }\right\}\left[ 1+\mathcal{O}\left(\frac{1}{M}\right)\right],
\label{C10}\end{equation}
\begin{equation}
Q^{(\sigma)} > Q^{(\sigma)}_{as} \left\{ { (1+\alpha)^{-r} \prod_{i=1}^r\left( \frac{X_i^{(+)}}{q_i}\right)^{-1} \atop 1-\alpha }\right\}\left[ 1+\mathcal{O}\left(\frac{1}{M}\right)\right],
\label{C11}\end{equation}
with
\begin{equation}
 Q^{(\pm)}_{as}  \equiv \exp\bigl\{ (N\pm M) K_r(\mathbf{X^{(\pm)}}|\mathbf{q})\bigr\}
\label{C12}\end{equation}
and fixed $q_i = K_i/M$ as $M\to \infty$. Eq. (\ref{C10}) will be of use in the proof of Eq. (\ref{E8}) in theorem 2.

%%%%%%%%%%%%%%%%%%%%%%%%%%%%%%%%%%%%%%%%%%%%%%%%%%%%%%%%%%

\end{document}